\documentstyle[seceq,mbf,amsfonts,amssymb]{ptptex}

\def\boldsymbol#1{{\mbf #1}}
\def\eqref#1{(\ref{#1})}

\def\text#1{\hbox{#1}}
\def\paragraph#1{\noindent{\bf #1}}

\def\linebreak{\hfill\break}

\def\therefore{\mbox{\setbox0=\hbox{X}\hbox{$\ldotp$}\raise0.7\ht0\hbox{$\ldotp$}\hbox{$\ldotp$}} \quad }
\def\because{\mbox{\setbox0=\hbox{X}\raise0.7\ht0\hbox{$\ldotp$}\hbox{$\ldotp$}\raise0.7\ht0\hbox{$\ldotp$}}\kern0pt }

\def\r#1{{\rm #1}}

\def\bm#1{\boldsymbol{#1}}

\def\Frac(#1/#2){\left(\frac{#1}{#2}\right)}

\def\Order#1{\r{O}\!\left(#1\right)}

\def\Eq#1{\begin{equation} #1 \end{equation}}
\def\Eqr#1{\begin{eqnarray} #1 \end{eqnarray}}

\def\Eqrsubl#1#2{\begin{subequations}\label{#1}
\Eqr{#2}\end{subequations}}
\def\Bitm{\begin{itemize}}
\def\Eitm{\end{itemize}}
\newtheorem{theorem}{Theorem}[section]

\def\ZR{{{\mathbb Z}}}

\def\OG{\hbox{\it O}}

\def\IO{\hbox{\it IO}}

\def\dS{\hbox{dS}}
\def\AdS{\hbox{AdS}}

\title{Rigidity theorems in the braneworld model}

\author{Hideo Kodama}

\inst{Yukawa Institute for Theoretical Physics\\
Kyoto University, Kyoto 606-8502, Japan}

\abst{In the present paper, we give some theorems representing  
ridigity of a vacuum brane in static bulk spacetimes. As an  
application, we show that a static bulk spacetime with dimension  
$D\ge 4$ and spatial symmetry $\IO(D-2)$, $\OG(D-1)$ or  
$\OG_+(D-2,1)$ does not allow a vacuum brane with a black hole on  
it. We also show that if a static bulk spacetime with dimension  
$D>4$ satsifying the vacuum Einstein equations can be foliated by a  
continuous family of vacuum branes with asymptotically constant  
curvature, it is a black string solution.
} 

\begin{document}

\maketitle

\section{Introduction}

Recently, braneworld models are actively studied as possible new  
universe models based on higher dimensional unified theories. In  
particular, for the Randall-Sundrum models,%
\cite{Randall.L&Sundrum1999,Randall.L&Sundrum1999a} 
detailed analyses have been done by many people, and it has been  
shown that local behavior of gravity is the same as the conventional  
one in low energies and FRW universe models can be implemented.%
\cite{Randall.L&Sundrum1999a,Garriga.J&Tanaka2000,%
Tanaka.T&Montes2000,Giddings.S&Katz&Randall2000,%
Kudoh.H&Tanaka2001,Vollick.D2001,Kraus.P1999,%
Cline.J&Grojean&Servant1999,Ida.D2000,%
Kodama.H&Ishibashi&Seto2000,Koyama.K&Soda2000,Kodama.H2000A,%
Langlois.D&&2001,Koyama.K&Soda2002} 
However, whether global behavior of gravity is consistent with  
observations is still unclear. In particular, the structure of black  
holes formed by gravitational collapse remains as an important  
problem to be clarified, in order to see the viability of  
braneworld models. 
 
In the present paper, we give some mathematical theorems concerning  
restrictions on spacetimes imposed by the existence of a vacuum  
brane, which may be relevant to investigations of black holes in  
brane world models. Here, by a vacuum brane, we mean a hypersurface  
whose extrinsic curvature $K_{\mu\nu}$ is a constant multiple of the  
induced metric $g_{\mu\nu}$: 
\Eq{
K_{\mu\nu}=\sigma g_{\mu\nu}. 
\label{UmbilicCondition}
}
This condition is equivalent to the condition that the  
energy-momentum tensor of the brane is proportional to its metric,  
if the bulk spacetime is empty and the $\ZR_2$ symmetry is imposed  
at the brane. We consider two situations. The first one is a brane  
in a static $D$-dimensional spacetime with spatial symmetry  
$G(D-2,K)$, where $G(n,K)$ is the isometry group of a  
$n$-dimensional space with constant curvature $K$. This is an  
extension of the problem considered in a paper by Chamblin, Hawking  
and Reall.%
\cite{Chamblin.A&Hawking&Reall2000}  
In this paper, they showed that a 5-dimensional Schwarzschild-Anti  
de Sitter spacetime does not contain a brane with a black hole,  
although there exists a 5-dimensional solution called the black  
string solution, which admits a brane slice with a Schwarzschild  
black hole in it. The second problem discussed in the present paper  
is related to this black string solution. We will show that this  
solution is characterized as a vacuum bulk solution to the Einstein  
equations that allows a slice by a continuous family of vacuum  
branes.  

\section{Brane in Static $D$-dimensional Spacetimes with Spatial  
Symmetry $G(D-2,K)$}
 
First, we consider a vacuum brane in a 
$D$-dimensional static bulk spacetime with spatial symmetry  
$G(D-2,K)$, whose metric is represented as
\Eq{
ds_D^2=-U(R)dT^2+\frac{dR^2}{W(R)}+S(R)^2 d\sigma_{D-2}^2,  
\label{bulkgeometry:symmetric}
}
where $d\sigma_{D-2}^2=\gamma_{ij}dz^i dz^j$ is the metric of a  
$(D-2)$-dimensional constant curvature space $M^{D-2}_K$ with  
sectional curvature $K$ ($K=0,\pm1$). We do not impose the Einstein  
equations. So, $U(R)$, $W(R)$ and $S(R)$ are arbitrary functions.  
Note that the $G(D-2,K)$ symmetry gives the bulk spacetime a natural  
bundle structure $(N^2,M^{D-2}_K)$, where the base space $N^2$ is a  
2-dimensional orbit spacetime with the coordinates $(T,R)$. 
 
Possible configurations of a brane in this spacetime are determined  
by the condition \eqref{UmbilicCondition}. When we expressed the  
brane configuration as, say, $R=R(T,\bm{z})$,  
\eqref{UmbilicCondition} gives a set of partial differential  
equations for $R(T,\bm{z})$. An interesting point here is that this  
set of equations do not have a solution for an arbitrary choice of  
$U(R)$, $W(R)$ and $S(R)$, and the consistency of the equations  
leads to strong restrictions on the bulk geometry. Since a detailed  
analysis of this problem is given in a separate paper by the  
author\cite{Kodama.H2002A}, we only give a brief summary of the  
results in the form of theorems here.  
 
First, for a brane with $\sigma=0$, the following theorem holds. 
 
\begin{theorem} 
Configurations of a brane with $\sigma=0$ and allowed geometries are  
classified into the following three types: 
\begin{itemize} 
\item[I-A)] Brane configurations that are represented by subbundles   
$\Sigma=(N^2,F)$ of $M^D=(N^2,M_K^{D-2})$, where $F$ is a totally  
geodesic hypersurface $M$. Configurations of this type always exist  
irrespective of the choice of $U,W$ and $S$, and are mutually  
isometric. Each configuration is invariant under $\IO(1)\times  
G(D-3,K')$ for some $K'\ge K$.  
\item[I-B)] $G(D-2,K)$-invariant configurations which are represented  
as $R=R(T)$ by solutions to 
\Eq{
R_T^2=WU(1-AU)\not=0;\ A>0.
}
The bulk geometry is restricted to a simple product $N^2\times  
M_K^{D-2}$, for which  $S=$constant. 
\item[I-C)] Static configurations expressed as $R=$constant in terms  
of solutions to 
\Eq{
U'=0,\quad S'=0.
}
Each configuration of this type corresponds to an $\IO(1)\times  
G(D-2,K)$-invariant totally geodesic hypersurface.  
\end{itemize} 
A brane with $\sigma=0$ can take only configurations of  
the type I-A in $\AdS^D$, while it takes configurations of the type  
I-B and I-C in $E^{D-1,1}$ and those of the type I-C in $\dS^D$.  
The latter are all mutually isometric in a given bulk geometry. In  
a bulk spacetime that does not have constant curvature,  
configurations of the type I-B or I-C become isometric to those of  
the type I-A only when the bulk spacetime has the product structure  
$E^{0,1}\times M_K^{D-1}$.  
\end{theorem} 
 
A brane with $\sigma=0$ does not play a physical role in most of  
braneworld models. However, since the condition $\sigma=0$ is  
equivalent to the condition that the hypersurface is totally  
geodesic, this theorem gives the complete classification of totally  
geodesic time-like hypersurfaces in spacetimes with the  
$\IO(1)\times G(D-2,K)$ symmetry and may be useful in other contexts. 
 
Next, for a brane with $\sigma\not=0$, the following theorem holds. 
 
\begin{theorem} 
A brane with $\sigma\not=0$ can exists only for special bulk geometries.  
These bulk geometries and corresponding brane configurations are classified  
into the following three types: 
\begin{itemize} 
\item[I-B)] Brane configurations that are $G(D-2,K)$-invariant and  
represented as $R=R(T)$ by solutions to  
\Eq{
R_T^2=U^2\left(1-\frac{U}{\sigma^2 R^2}\right)\not\equiv0.
}
The bulk geometry is restricted to those with $U=W$ and $S=R$. For  
$S=R$, the condition $U=W$ is equivalent to the condition that the  
Ricci tensor is isotropic in planes orthogonal to  
$G(D-2,K)$-orbits. \item[I-C)] Static and $G(D-2,K)$-invariant  
brane configurations expressed as $R=$constant in terms of  
solutions to 
\Eq{
\frac{U'}{U}=\frac{2S'}{S},\quad
W\left(\frac{S'}{S}\right)^2=\sigma^2.
\label{TypeI-C:condition}}
\item[III)] Static brane configurations in the bulk geometries with metrics  
of the form 
\Eq{ 
ds_D^2=d\sigma_{\lambda,D-1}^2 - UdT^2,
\label{TypeIII:metric}}
where $d\sigma_{\lambda,D-1}^2$ is the metric of a 
$(D-1)$-dimensional constant curvature space 
$M_\lambda^{D-1}$ with sectional curvature $\lambda$. $U$ is a  
function on this space that is invariant under a subgroup $G(D-2,K)$  
of the isometry group of $M_\lambda^{D-1}$. Allowed forms of $U$ and  
brane configurations are given as follows: 
\begin{itemize} 
\item[i)] $\lambda=0$: In a Cartesian coordinate system $\bm{x}$ for  
$E^{D-1}$, $U=((\bm{x}-\bm{a})^2+k)^2$ and brane configurations are   
represented as $(\bm{x}-\bm{b})^2=1/\sigma^2$ with  
$(\bm{b}-\bm{a})^2=1/\sigma^2-k$. 
\item[ii)] $\lambda=1/\ell^2>0$: In a homogeneous coordinate system  
$X$ in which $S^{D-1}$ is expressed as $X\cdot X=1$, $U=(P\cdot  
X+k)^2$ ($k\not=0$) and brane configurations are represented as  
$Q\cdot X=1$ with $Q\cdot P=-k$ and $Q\cdot Q=1+1/(\ell\sigma)^2$. 
\item[iii)] $\lambda=-1/\ell^2<0$: In a homogeneous coordinate  
system $Y$ in which $H^{D-1}$ is expressed as $Y\cdot Y=-1$,  
$U=(P\cdot Y +k)^2$ ($k\not=0$) and brane configurations are  
represented as $Q\cdot Y=1$ with $Q\cdot P=k$ and $Q\cdot  
Q=1/(\ell\sigma)^2-1$.  
\end{itemize} 
These brane configurations are not $G(D-2,K)$-invariant, but  
$G(D-3,K')$-invariant for some $K'\ge K$ and mutually isometric. 
\end{itemize}
\end{theorem}
 
These results implies that a vacuum brane in static $D$-dimensional  
spacetime with spatial symmetry $G(D-2,K)$ cannot contains a black  
hole. It is because if a brain contains a black hole, it does not  
belong to the type I, and hence the bulk geometry must be of the  
type III with special forms of $U$ given above. However, for these  
bulk geometries, the curvature diverges at points where $U$  
vanishes. Hence, it has no horizon.

\section{Static Spacetimes with Brane Slices}
 
Next, we consider a static configuration of a vacuum brane $\Sigma$  
in a $D$-dimensional static bulk spacetime $M^D$. Now, we assume  
that the bulk geometry is a solution to the vacuum Einstein  
equations with cosmological constant $\tilde \Lambda$. However, we  
do not assume any spatial symmetry of the bulk geometry. 
 
In an appropriate coordinate system $(y,x^\mu)$ in which $\Sigma$ is  
represented by the hypersurface $y=0$, the metric of the bulk  
spacetime can be written 
\Eq{ 
ds_D^2= N(y,x)^2 dy^2+g_{\mu\nu}(y,x)dx^\mu dx^\nu. 
\label{metric:braneslice}} 
In this coordinate system, the extrinsic curvature of each  
$y=$constant surface $\Sigma_y$ is expressed as  
\Eq{ 
\partial_y g_{\mu\nu}=2NK_{\mu\nu}. 
\label{ExtrinsicCurvature:def}} 
The Einstein equations provide a kind of evolution equations for  
this extrinsic curvature, 
\Eqrsubl{Einstein:dotK}{ 
&& \partial_y K + \Box N + (K^2-R)N=-\frac{2(D-1)}{D-2}\tilde\Lambda  
N, 
\label{Einstein:dotK:trace}\\ 
&& \partial_y \hat K^\mu_\nu + NK\hat K^\mu_\nu 
+\nabla^\mu\nabla_\nu N -\frac{1}{D-1}\Box N  
\delta^\mu_\nu=NS^\mu_\nu, 
\label{Einstein:dotK:traceless}, 
} 
and the constraints 
\Eqrsubl{Einstein:Constraints}{ 
&& -R+\frac{D-2}{D-1}K^2-\hat K^\mu_\nu \hat  
K^\nu_\mu=-2\tilde\Lambda, 
\label{Einstein:HamiltonianConstraint}\\ 
&& \nabla_\nu \hat K^\nu_\mu-\frac{D-2}{D-1}\partial_\mu K=0, 
} 
where $K=K^\mu_\mu$, $\hat K^\mu_\nu$ is the tracefree part of  
$K^\mu_\nu$, $\nabla_\mu$ is the covariant derivative with respect  
to $g_{\mu\nu}$ on $\Sigma_y$, and $S^\mu_\nu$ is the tracefree part  
of the Ricci curvature $R^\mu_\nu$ of $g_{\mu\nu}$. 
 
If we assume that $N$ depends only on $y$ and that $g_{\mu\nu}$ has  
the product form $e^{2\Phi(y)}\hat g_{\mu\nu}(x)$, after  
reparametrizing $y$ so that $N=1$, the extrinsic curvature is  
written in the form \eqref{UmbilicCondition} with $\sigma=\partial_y  
\Phi(y)$. Hence, the evolution equations for $K_{\mu\nu}$ reduce to  
\Eqrsubl{BS:dot K}{ 
&& S^\mu_\nu\equiv R^\mu_\nu -\frac{R}{D-1}\delta^\mu_\nu=0,\\ 
&& \partial_y \sigma+(D-1)^2\sigma^2-R 
=-\frac{2(D-1)}{D-2}\tilde\Lambda, 
} 
and the constraints give 
\Eq{ 
R=(D-1)(D-2)\sigma^2+2\tilde\Lambda. 
\label{Constraint:Umbilic}} 
Since the last equation implies that $R$ depends only on $y$, it  
follows that each $\Sigma_y$ is an Einstein spacetime, and that  
when an Einstein metric on $\Sigma=\Sigma_0$ is given, the structure  
of the bulks spacetime, including $\Phi(y)$, is uniquely  
determined. This is the well-know black string solution.%
\cite{Chamblin.A&Hawking&Reall2000,Garriga.J&Sasaki2000}

In this section, we show that this black string solution can be  
characterized as a spacetime in which a vacuum brane can be  
non-isometrically deformed continuously, i.e., is not rigid. In the  
present paper, we only consider the case in which the brane  
configurations as well as the bulk spacetime are static. We assume  
that $D>4$. Hence, the metric of the brane can be written 
\Eq{ 
ds_\Sigma^2=-V(\bm{x})^2dt^2+g_{ij}(\bm{x})dx^idx^j, 
} 
and $N$ is independent of $t$. We also assume that $N$ and $D_iN$ is  
bounded on $\Sigma$, where $D_i$ is the covariant derivative with  
respect to $g_{ij}$. The basic equations are  
\eqref{ExtrinsicCurvature:def}, \eqref{Einstein:dotK} and  
\eqref{Einstein:Constraints}, which hold also for an infinitesimal  
deformation of $\Sigma$ if $\delta y N$ is regarded as a   
displacement of $\Sigma$ by the deformation along the unit normal  
to $\Sigma$.

Now, let us assume that  
\Eq{ 
\partial_y \hat K^\mu_\nu=0 
\label{UmbilicDeformation} 
} 
holds in addition to \eqref{UmbilicCondition} on $\Sigma$. Then, the  
Einstein equations reduce to \eqref{Constraint:Umbilic} and the set  
of equations, 
\Eqrsubl{UmbilicDeformationEqs}{ 
&& V\Box N\equiv D\cdot (VDN)=-\tilde N V, 
\label{UDE:BoxN}\\ 
&& V\nabla^t\nabla_t DV\cdot DN=-\frac{\tilde N}{D-1}V+VNS^t_t, 
\label{UDE:DtDtN}\\ 
&& D_iD_j N=-\frac{\tilde N}{D-1}g_{ij}+ NS_{ij}, 
\label{UDE:DiDjN} 
} 
where 
\Eq{ 
\tilde N=\frac{R}{D-2}N+\partial_y K,
} 
and $\sigma, \partial_y K$ and $R$ are required to be constant.   
 
First, we consider the case $R=0$. In this case, we further assume  
that the geometry of the brane is asymptotically flat and  
$S^\mu_\nu$ falls off sufficiently rapidly: 
\Eqr{ 
&& g_{ij}dx^idx^j=-V^{-2}dr^2 
+r^2\hat \gamma_{AB}d\theta^Ad\theta^B+\Order{\frac{dx\cdot  
dx}{r^{D-3}}},\\ 
&& V=1-\frac{2\mu}{r^{D-4}}+\Order{\frac{1}{r^{D-3}}},\quad 
S^\mu_\nu={\rm o}\left(\frac{1}{r^{D-2}}\right). 
} 
Then, from \eqref{UDE:DtDtN} it follows that $\partial_y K$  
vanishes. Further, \eqref{UDE:DiDjN} requires that $N$ behaves as $N  
=N_0+{\rm o}(1/r^{D-4})$, where $N_0$ is some constant. In the  
meanwhile, multiplying \eqref{UDE:BoxN} by $N$ and integrating it  
over a $t=$constant hypersurface $F$ in $\Sigma$, we obtain 
\Eq{ 
\int_F d^{D-2}x \sqrt{g}VD^iN D_iN=\int_{\partial F}dS_i VND^iN,} 
where $\partial F$ consists of the sphere at $r=\infty$ and horizons  
where $V$ vanishes. It is easy to see that these boundary  
contributions vanish from the above asymptotic estimate and the  
regularity of $N$ and $D^iVD_iN$ deduced from the above basic  
equations. This implies that $N$ is constant outside the horizons  
and the metric $g_{\mu\nu}$ of the brane is a solution to the vacuum  
Einstein equation $R_{\mu\nu}=0$. 
 
Next, let us consider the case $R=-(D-1)(D-2)/\ell^2<0$. In this  
case, we assume that the brane geometry is asymptotically anti-de  
Sitter in the sense 
\Eqr{ 
&& g_{ij}dx^idx^j=\frac{dr^2}{V^2+\Order{1/r^{D-3}}} 
+r^2\left(\hat \gamma_{AB}+\Order{\frac{1}{r}}\right)d\theta^A  
d\theta^B \nonumber\\ 
&& \qquad\qquad +\Order{\frac{1}{r^{D-4}}}dr d\theta^A,\\ 
&& V^2=1+\frac{r^2}{\ell^2}-\frac{2\mu}{r^{D-4}} 
+\Order{\frac{1}{r^{D-3}}},\quad 
S^\mu_\nu={\rm o}\left(\frac{1}{r^{D-3}}\right). 
} 
Under this condition, from \eqref{UDE:DiDjN} it follows that $\tilde  
N$ falls off as $o(1/r^{D-3})$ at infinity. Then, by a similar  
argument as that in the case $R=0$, we find that \eqref{UDE:BoxN}  
requires that $\tilde N$ vanishes. Hence, the brane geometry must be  
Einstein again. 
 
Unfortunately, we cannot show that $\tilde N$ vanishes by the same  
method for the case $R>0$, because in the equation $D^i(\tilde N  
VD_i\tilde N)=VD^i\tilde N D_i\tilde N-RV\tilde N^2/(D-2)$ obtained  
from \eqref{UDE:BoxN}, the right-hand side does not have a definite  
sign. However, under a stronger assumption that $N$ does not vanish  
outside horizons, we can obtain the same conclusion by a different  
method. The basic idea is to use the identity  
\Eq{ 
\int_{A} d^{D-1}\sqrt{g} NS^{\mu\nu}S_{\mu\nu} 
=\int_{\partial A}d\Sigma^\mu S_{\mu\nu}\partial^\nu N 
} 
obtained from \eqref{UmbilicDeformationEqs}, where $A$ is a region  
outside horizons in $\Sigma$, cut out by two $t=$constant  
hypersurfaces. We assume that the region of $\Sigma$ outside the  
horizons are spatially compact, because the brane is de Sitter like.  
Then, we can easily show that the contributions from the horizons on  
the right-hand side of this equation vanish. Further, the  
contributions from the $t=$constant hypersurfaces cancel due to the  
staticity. Hence, the left-hand side of this equation vanishes. If  
we take into account the assumption on $N$ and the fact that  
$S^\mu_\nu S^\nu_\mu=(S^t_t)^2+S_{ij}S^{ij}\ge0$, this implies that  
$S_{\mu\nu}$ vanishes and $\nabla_\mu \nabla_\nu N=cg_{\mu\nu}$,  
where $c$ is a constant. Then, applying the Bianchi identity to the  
divergence of the latter equation, we obtain $\nabla_\mu N=0$.  
Hence, we find that $N$ is constant and the brane geometry is  
Einstein.  
 
The results obtained in this section are summarized as follows. 
 
\begin{theorem} 
Let $\Sigma$ be a static vacuum brane $\Sigma$ in a static  
$D$-dimensional spacetime with $D>4$ satisfying the Einstein  
equations with $\tilde \Lambda$, and consider a static infinitesimal  
deformation of $\Sigma$ along its normal proportional to a function  
$N$ on $\Sigma$.  Then, if the deformation preserves the isotropy of  
the extrinsic curvature, the Ricci scalar $R$ of $\Sigma$ is  
constant. Further, if $\Sigma$ is asymptotically flat ($R=0$) or  
asymptotically $\AdS$ ($R<0$) and $N$ is uniformly bounded, $\Sigma$  
is an Einstein spacetime and $N$ is a constant. The same result  
holds also for $R>0$, if $N$ does not vanish outside horizons on  
$\Sigma$. 
\end{theorem} 
 
If the bulk spacetime is analytic, it immediately follows from this  
theorem that an infinitesimal deformation of the brane with the  
required property exists only when the geometry of the bulk  
spacetime is of the black string type, because the bulk geometry is  
uniquely determined by the metric and the extrinsic curvature on the  
brane. Further, even when the analyticity is not assumed, we can  
obtain the same conclusion if we require that the bulk spacetime can  
be foliated by a family of vacuum branes with bounded $N$. It is  
because the theorem guarantees that the assumption used to obtain  
the black string solution holds in this case, if we choose the  
$y$-coordinate in \eqref{metric:braneslice} so that $y$ is constant  
on each brane. Hence, the following theorem holds. 
 
\begin{theorem} 
If the bulk spacetime with dimension $D>4$ satisfying the vacuum  
Einstein equations can be foliated by a family of vacuum branes  
with bounded $N$ and asymptotically constant curvature, it is a  
black string solution. 
\end{theorem}


\end{document}